\documentclass{emulateapj}
\usepackage{natbib}
\usepackage{color}
\newcommand{\tpeak}{\ensuremath{\tau_\mathrm{peak}}}
\newcommand{\tcen}{\ensuremath{\tau_\mathrm{cen}}}
\newcommand{\rblr}{\ensuremath{R_{\mathrm{BLR}}}}
\newcommand{\mbh}{\ensuremath{M_\mathrm{BH}}}
\newcommand{\fvar}{\ensuremath{F_\mathrm{var}}}

\slugcomment{ApJL, accepted}
\shorttitle{Robotic Reverberation Mapping}
\shortauthors{Valenti et al.}

\begin{document}
 \title{Robotic Reverberation Mapping of Arp~151}

\author{S. Valenti,$\!$\altaffilmark{1,2} D.J. Sand,$\!$\altaffilmark{3} A. J. Barth,$\!$\altaffilmark{4} 
 K. Horne,$\!$\altaffilmark{5}  T. Treu,$\!$\altaffilmark{6,2} 
 L. Raganit,$\!$\altaffilmark{3}  T. Boroson,$\!$\altaffilmark{1}    S.~Crawford,$\!$\altaffilmark{7}
 \and
 A. Pancoast,$\!$\altaffilmark{8,9} L. Pei,$\!$\altaffilmark{4}, 
E. Romero-Colmenero,$\!$\altaffilmark{7,10} C. Villforth,$\!$\altaffilmark{5,11}   H. Winkler,$\!$\altaffilmark{12} 
 } 

\begin{abstract}
We present the  first results from the Las  Cumbres Observatory Global
Telescope  (LCOGT) Network's  Active  Galactic Nuclei  Key Project,  a
large  program devoted  to using  the  robotic resources  of LCOGT  to
perform  time domain  studies of  active galaxies.   We monitored  the
Seyfert  1 galaxy  Arp~151 (Mrk~40)  for $\sim$200  days with  robotic
imagers and with the FLOYDS  robotic spectrograph at Faulkes Telescope
North.   Arp~151  was  highly  variable  during  this  campaign,  with
$V$-band light  curve variations  of $\sim$0.3  mag and  H$\beta$ flux
changing by a factor of $\sim$3.   We measure robust time lags between
the  $V$-band  continuum and  the  H$\alpha$,  H$\beta$ and  H$\gamma$
emission    lines,     with    $\tcen     =    13.89^{+1.39}_{-1.41}$,
7.52$^{+1.43}_{-1.06}$ and  7.40$^{+1.50}_{-1.32}$ days, respectively.
The lag for the \ion{He}{2} $\lambda4686$ emission line is unresolved.
We measure  a velocity-resolved  lag for the  H$\beta$ line,  which is
clearly asymmetric with higher lags on the blue wing of the line which
decline  to the  red, possibly  indicative  of radial  inflow, and  is
similar in  morphology to past  observations of the  H$\beta$ transfer
function  shape.   Assuming  a  virialization factor  of  $f$=5.5,  we
estimate         a         black          hole         mass         of
$\mbh=6.2^{+1.4}_{-1.2}\times$10$^{6}$~$M_{\odot}$,   also  consistent
with past measurements for this object.  
These results represent the first step to demonstrate the powerful 
robotic capabilities of LCOGT for long-term, AGN time domain 
campaigns that human intensive programs cannot easily accomplish.
Arp 151 is now  one of
just a  few AGN where the  virial product is known  to remain constant
against substantial changes in H$\beta$ lag and luminosity.
\end{abstract}

 
\altaffiltext{1}{Las Cumbres Observatory Global Telescope Network, 6740 Cortona Dr., Suite 102, Goleta, CA 93117, USA}
\altaffiltext{2}{Department of Physics, University of California, Santa Barbara, Broida Hall, Mail Code 9530, Santa Barbara, CA 93106-9530, USA}
\altaffiltext{3}{Texas Tech University, Physics Department, Box 41051, Lubbock, TX 79409-1051, USA}
\altaffiltext{4}{Department of Physics and Astronomy, 4129 Frederick Reines Hall, University of California, Irvine, CA, 92697-4575, USA}
\altaffiltext{5}{SUPA School of Physics \& Astronomy, North Haugh, Univ. of St.Andrews, KY16~9SS, Scotland, UK.} 
\altaffiltext{6}{Department of Physics and Astronomy, University of California, Los Angeles, CA, USA 90095-1547}
\altaffiltext{7}{South African Astronomical Observatory, PO Box 9, Observatory 7935, Cape Town, South Africa}
\altaffiltext{8}{Harvard-Smithsonian Center for Astrophysics 60 Garden St Cambridge, MA 02138 USA}
\altaffiltext{9}{Einstein Fellow}
\altaffiltext{10}{Southern African Large Telescope Foundation, PO Box 9, Observatory 7935, Cape Town, South Africa}
\altaffiltext{11}{University of Bath, Department of Physics, Claverton Down Rd, Bath, BA2 7AY, UK}
\altaffiltext{12}{Department of Physics, University of Johannesburg, PO Box 524, 2006 Auckland Park, South Africa}

\section{Introduction}

Measuring  supermassive black  hole  masses (\mbh)  as  a function  of
galaxy  mass  and cosmic  time  is  one  of  the great  challenges  in
observational astrophysics.  A vital technique for  measuring \mbh\ is
reverberation   mapping   of   Type    1,   active   galactic   nuclei
\citep[AGN;][]{blandford82,Peterson93}.  Reverberation  mapping traces
the time  lag between variable  continuum emission from  the accretion
disk and the  larger gas clouds that this  emission photoionizes. This
results  in  broad  emission   line  variability  lagging  behind  the
continuum emission.  The lag time provides a distance scale due to the
light  travel time  between the  continuum source  and the  broad line
region (BLR), while the broad  line Doppler width provides a velocity.
From these, and  assumptions about the geometry and  kinematics of the
BLR,  one can  infer  $M_{\rm BH}$.   Reverberation  mapping has  been
applied         to         dozens        of         nearby         AGN
\citep[e.g.][]{Peterson04,bentz09,bentz10lags,denney10,grier12,Du14},
and the resulting $\rblr -  L$ relation \citep[e.g.][]{bentz13} is the
basis for the  ``single-epoch'' virial mass methods  employed to track
the growth history and redshift  evolution of supermassive black holes
\citep[for a recent review see][]{Shen13}.

Reverberation mapping  campaigns are labor intensive  and expensive in
terms  of  telescope  time  and  human  effort.   They  often  require
telescope  time allocations  over  multiple  periods with  classically
scheduled facilities.

A potentially  transformative facility for reverberation  mapping, and
time domain studies,  is the Las Cumbres  Observatory Global Telescope
network  \citep[LCOGT;][]{Brown13}.  LCOGT  is  currently composed  of
nine 1-m  imaging telescopes  at sites  around the  globe and  two 2-m
Faulkes  telescopes,  equipped with  imagers  and  the FLOYDS  robotic
spectrographs.  LCOGT  is completely  robotic, from the  scheduling of
the  telescopes   to  the  placement  of   targets  onto  spectrograph
slits.  Facilities  such  as  LCOGT  are  ideally  suited  to  conduct
intensive  reverberation  mapping  campaigns, as  well  as  persistent
multi-season efforts on luminous or higher redshift AGN.

The AGN  Key Project  (PI: K.  Horne)  has been  allocated substantial
resources  on  the  LCOGT  network to  conduct  several  AGN  studies,
including  reverberation  mapping  of  both  local  and  high-redshift
objects. Here  we report our  first results -- an  intensive $\sim$200
day  long reverberation  mapping campaign  of the  well-studied, local
Seyfert 1 galaxy Arp~151 (also known as Mrk 40).  Arp~151 was selected
as part  of an initial sample  of low-redshift AGNs at  LCOGT based on
its  strong variability  in past  campaigns \citep{Bentz08,  Barth15}.
Our   all-robotic   observations   show  a   clear   velocity-resolved
reverberation lag.

These data  allow us to  investigate the  time variability of  the BLR
over a time much longer than the  lag and to infer the black hole mass
in Arp 151.  We verify the virial product remains constant independent
of  an AGN  luminosity  change,  and demonstrate  the  promise of  AGN
studies with the LCOGT network.

\section{Photometry}\label{sec:imaging}

We obtained  high signal  to noise ratio  $V$-band imaging  of Arp~151
from  11  November  2014  to  4 July  2015  with  near-daily  cadence.
Observations  were carried  out  with the  1-m  telescope at  McDonald
Observatory and  the 2-m Faulkes  Telescope North (FTN)  at Haleakala,
Hawaii.   The  McDonald  1m  telescope  was fitted  out  with  a  SBIG
STX-16803 CCD with a 15\farcm8$\times$15\farcm8 field of view, sampled
with  0\farcs464 pixels.   The FTN  observations were  taken with  the
Spectral camera, with a 10\farcm5$\times$10\farcm5 field of view and a
binned  pixel  scale  of   0\farcs30.   Typical  exposure  times  were
2$\times$120 s per visit.

Initial photometric data reduction, including overscan subtraction and
flat fielding, was  done via the LCOGT  data pipeline \citep{Brown13}.
After this,  the images  were cleaned  of cosmic  rays using  the L.A.
Cosmic  algorithm  \citep{vandokkum01}.   In  order  to  register  the
celestial  coordinates across  the set  of  images, we  used the  {\it
  Astrometry.net} software package \citep{Lang10}.

Aperture  photometry was  applied using  the automated  IDL procedures
described by \citet{Pei14}, using four comparison stars to construct a
relative light curve.  An aperture radius of 5\arcsec\ and sky annulus
radii  of  15\arcsec--30\arcsec\  were  used  throughout.   Photometry
measurements  from images  taken within  6  hours of  each other  were
combined to produce  a weighted average magnitude for  the final light
curve. No  attempt was made to  remove host-galaxy light from  the AGN
photometry.   We  then  used  the  AAVSO  Photometric  All-Sky  Survey
\citep[APASS;][]{Henden12}  to flux  calibrate  the  data.  The  final
$V$-band light  curve of 119  data points  exhibits a sharp  flux peak
with  amplitude $\sim$0.3  mag,  ideal for  measurement of  broad-line
reverberation (Figure~\ref{fig:lcs}).

\section{Spectroscopic Observations and Reduction}\label{sec:spec}

FLOYDS robotic spectroscopic observations  of Arp~151 were carried out
between 2014 December 6 and 2015  June 5.  All observations were taken
at the parallactic angle with a 1\farcs6$\times$30\arcsec\ slit and an
exposure time of 3600 s.  In  all, 55 spectra of Arp~151 contribute to
our emission line light curve analysis.

The FLOYDS spectrographs are a pair of nearly identical instruments at
the Faulkes Telescopes North and South (Arp~151 spectra, however, were
all observed with FTN).  They operate totally robotically, drastically
lowering  the human  effort necessary  to conduct  long and  intensive
reverberation mapping campaigns.  Targets  are automatically placed in
the  slit using  an acquisition  camera and  a fast  cross correlation
algorithm to recognize  the field and place the science  object in the
slit.  FLOYDS uses a low  dispersion grating (235 lines mm$^{-1}$) and
a  cross-dispersed   prism  to   work  in   first  and   second  order
simultaneously, giving a single-shot wavelength coverage of $\sim$3200
to 10000.  First order wavelength  coverage extends over $\sim$4800 to
10000 \AA\  with a pixel  scale of 3.51 \AA\  pixel$^{-1}$ (resolution
13\AA),  while  second  order  covers   3200  to  5900  \AA\  at  1.74
\AA\ pixel$^{-1}$ (resolution 7\AA) --  for convenience, we will refer
to  these as  the red  and blue  side, respectively.   The significant
order overlap allows the H$\beta$ emission line to be observed in both
orders   simultaneously,  providing   a   check   on  our   systematic
uncertainties  related  to  emission line  measurements.   The  median
continuum S/N per pixel is 22.5 at 4600 \AA\ on the blue side and 61.2
at 6400 \AA\ on the red side.

The spectroscopic data  were batch reduced with a  modified version of
the FLOYDS data reduction pipeline  called {\sc AGNfloyds}, written in
a PyRAF/Python framework.  Briefly,  {\sc AGNfloyds} performs standard
image detrending (overscan subtraction, flat fielding and defringing),
cosmic     ray    rejection     \citep[via     the    L.A.      Cosmic
  algorithm;][]{vandokkum01},     order    rectification,     spectral
extraction, and flux and  wavelength calibration.  Each spectral order
is processed  separately.  We  utilize a fixed,  unweighted extraction
aperture  of  8\farcs8  to  minimize  the  effect  of  varying  seeing
conditions throughout  the campaign.  Error spectra  were produced and
propagated  forward for  the emission  line light  curve measurements.
All spectra were reduced without human intervention.

\section{Emission line light curves}

Emission line measurements were carried out on the FLOYDS spectra in a
similar  way   as  the  recent   Lick  AGN  Monitoring   Project  2011
reverberation mapping campaign  \citep{Barth11,Barth15}, with the blue
and red side spectra treated separately.
  
\begin{figure*}
\begin{center}
\mbox{ \epsfysize=10.0cm \epsfbox{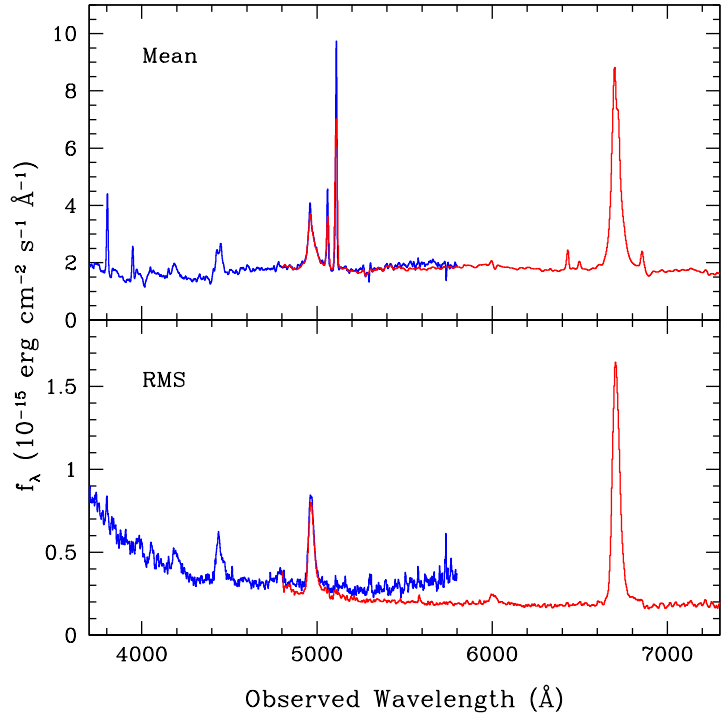}} 
\caption{Mean and  root mean  square spectra  of Arp~151  from FLOYDS,
  displaying both the blue and red side spectra. Note that H$\beta$ is
  in  the wavelength  overlap region  of FLOYDS,  and appears  in both
  first and  second order.  A  small flux  scaling was applied  to the
  red-side spectrum to match that of the blue side.  The upturn in the
  red side  of the blue rms  spectrum is due  to the lower S/N  of the
  data as  the second  order throughput of  the grating  declines. 
\label{fig:meanrms} }
\end{center}
\end{figure*}

First,  the  reduced  spectra  were  placed  on  a  uniform  flux  and
wavelength  scale  utilizing  the  procedure  of  \citet{vG92},  which
applies a  linear wavelength shift,  flux scaling factor  and Gaussian
broadening term to minimize differences  in the spectral region around
the [\ion{O}{3}]  $\lambda$5007 line  from night  to night,  under the
assumption  that this  narrow  line  has a  constant  flux during  the
spectroscopic  campaign.   After  this procedure  was  performed,  the
excess  rms   scatter  of   the  [\ion{O}{3}]   light  curve   in  our
spectroscopic time series was $\sim$2\% for both the blue and red side
spectra, similar  to recent reverberation  mapping work, which  can be
considered as a measure of the random error in flux calibration of the
final scaled spectra \citep[e.g.][]{Barth15}.   The final mean and rms
spectra can be seen in Figure~\ref{fig:meanrms}, constructed following
\citet{Peterson04}; note that the  [\ion{O}{3}] line almost completely
disappears in the  rms spectrum of both the red  and blue arm, showing
the high quality of the dataset.

For the blue-side scaled  spectra, we performed spectral decomposition
over a wavelength range of  $\sim$4200--5300 \AA~in a manner identical
to \citet{Barth15}  in order  to isolate  the H$\beta$,  H$\gamma$ and
\ion{He}{2} emission lines for our light curve measurements.  Briefly,
an eight-component model is fit to the daily, scaled spectra utilizing
the IDL  {\sc mpfit}  package \citep{Markwardt09}.  The  components of
the model include: 1) a stellar spectrum with a single burst age of 11
Gyr  and  solar  metallicity  \citep{Bruzual03}; 2)  a  power-law  AGN
featureless  continuum component;  3) the  [\ion{O}{3}] $\lambda$4959,
5007  lines;  4)  broad  and  narrow H$\beta$;  5)  broad  and  narrow
\ion{He}{2};   6)  broad   \ion{He}{1};   7)   an  \ion{Fe}{2}   blend
contribution  utilizing  the  templates   of  \citet{K10};  and  8)  a
\citet{Cardelli89} reddening law with the  color excess, $E(B-V)$ as a
free  parameter.   The  wavelength  region around  the  H$\gamma$  and
[\ion{O}{3}] $\lambda$4363  lines was masked out  from 4280--4400 \AA~
during  the fitting  process due  to their  complexity \citep[see][for
  details]{Barth15}.

\begin{figure*}
\begin{center}
\mbox{ \epsfysize=15.0cm \epsfbox{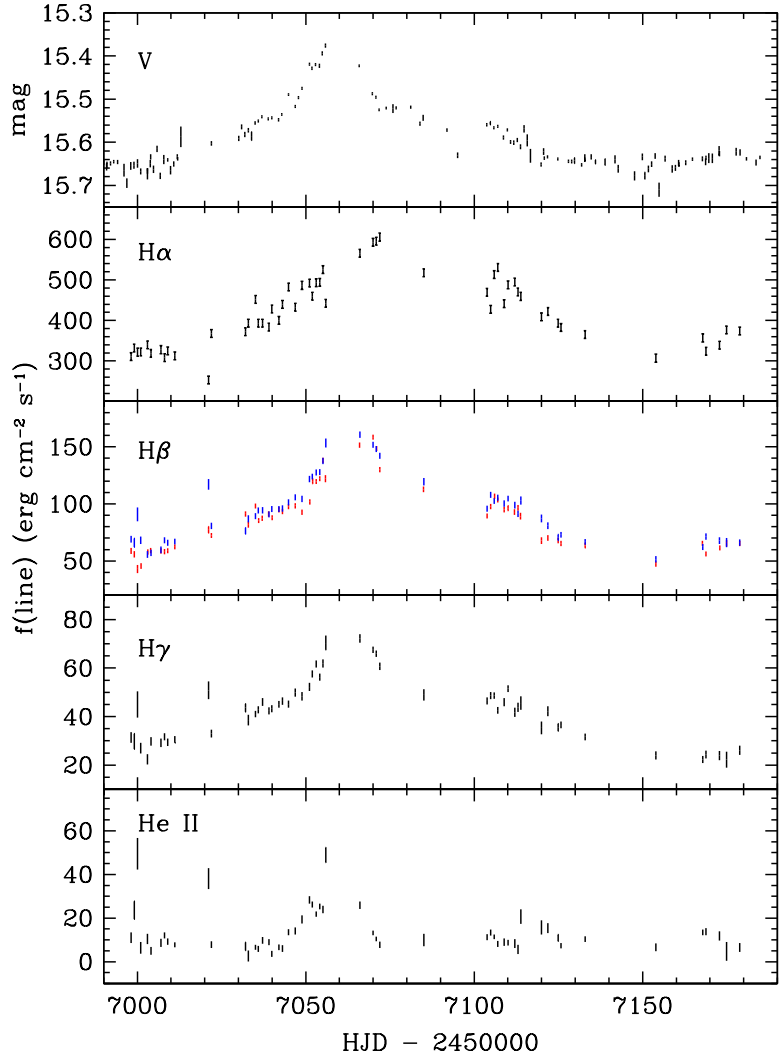}} 
\caption{ Arp~151 light curves for the $V$-band and the emission lines
  H$\alpha$, H$\beta$,  H$\gamma$, and \ion{He}{2} as  determined from
  the FLOYDS spectra.The  red and blue points show  the H$\beta$ light
  curve  determined  from  the  1st  and  2nd  order  FLOYDS  spectra,
  respectively,   as   this  emission   line   is   visible  in   both
  simultaneously. 
The normalized excess rms scatter of the [OIII] light curve has 
been added in quadrature to the photon-counting error bars on each 
data point to take in account of systematic errors.
  \label{fig:lcs}}
\end{center}
\end{figure*}

Once the best-fitting spectral  decomposition is found, all components
except for  the broad  and narrow  H$\beta$ component  were subtracted
from each spectrum, leaving only  the H$\beta$ and H$\gamma$ lines and
fitting residuals  in the  resulting spectra.   Light curves  of these
decomposed spectra were  generated via direct integration,  and can be
seen in  Figure~\ref{fig:lcs}.  Note that  both of these  light curves
include  a  narrow  line  component, and  the  H$\gamma$  light  curve
includes a contribution from the [\ion{O}{3}] $\lambda$4363 line.  The
light curve of the relatively  weak \ion{He}{2} $\lambda$4686 line was
determined  directly from  each night's  best-fitting model  component
spectrum,  which  includes  both  broad and  narrow  components.   The
\ion{Fe}{2} emission  proved to be  too weak  to yield a  useful light
curve.

For  the red-side  scaled spectra,  we measure  both the  H$\beta$ and
H$\alpha$ (which  includes contributions from the  narrow [\ion{N}{2}]
$\lambda$6548,  6583   lines)  fluxes  by  direct   integration  after
subtracting a simple linear continuum from adjacent line-free regions.
The H$\beta$  light curve derived  in this way agrees  remarkably well
with  the  FLOYDS blue-side  H$\beta$  light  curve, derived  via  our
spectral fits (compare  the blue and red points in  the H$\beta$ panel
of Figure~\ref{fig:lcs}).

The  emission  line  light   curves  in  Figure~\ref{fig:lcs}  show  a
morphology similar  to that seen in  the $V$-band light curve,  with a
factor of  3 rise and decline  during the $\sim$200 day  period of our
campaign. For all of our  emission-line light curves, we have measured
the         normalized        rms         variability        amplitude
\fvar\  \citep[e.g.,][]{Peterson04}.   We   present  these  values  in
Table~\ref{table:cc_results}.   In  general,   values  of  $\fvar>0.1$
correspond to strong line variations that are amenable to accurate lag
measurement  (provided  that  the   light  curve  exhibits  sufficient
structure  rather than  a  simple monotonic  increase  or decrease  in
flux).   The  light curves  for  H$\alpha$,  H$\beta$, H$\gamma$,  and
\ion{He}{2} all exhibit variations well in excess of this threshold.

We also  note that  as a  consistency check  we derived  emission line
light curves using the software  suite {\sc prepspec}, developed by K.
Horne \citep[see, e.g.][for a  description of the software]{grier13a},
and found similar light curves for all the lines measured here.

\section{Reverberation lag measurements} \label{sec:lag}

\begin{figure*}
\begin{center}
\mbox{ \epsfysize=10.0cm \epsfbox{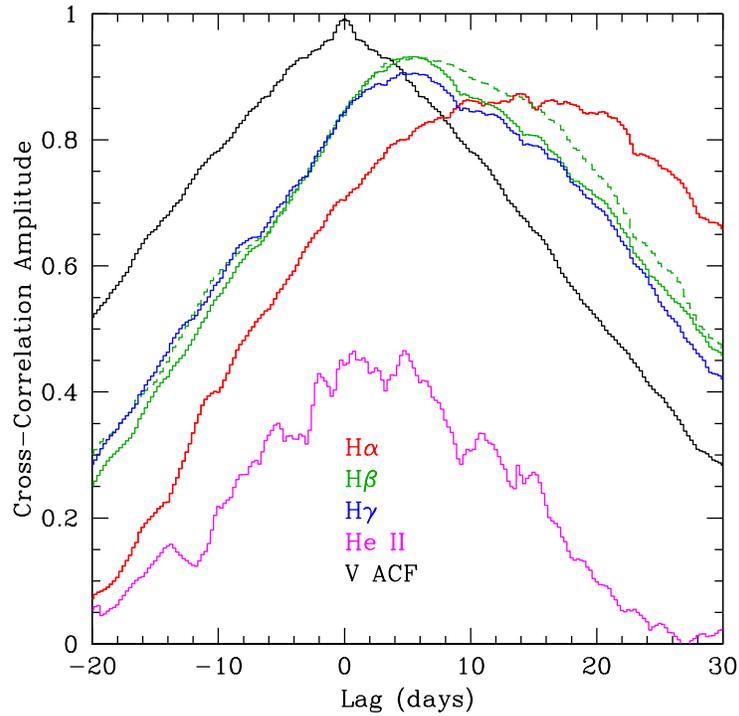}} 
\caption{ Cross  correlation functions  of H$\alpha$,  H$\beta$ (solid
  line  is the  blue side  data, dashed  line is  from the  red side),
  H$\gamma$ and  \ion{He}{2} against  the $V$-band light  curve.  Also
  shown   is   the   autocorrelation    function   of   the   $V$-band
  continuum. \label{fig:ccf}}
\end{center}
\end{figure*}

We  employ  standard   cross-correlation  function  (CCF)  techniques,
utilizing the interpolation and Monte Carlo error analysis methodology
of        previous         reverberation        mapping        workers
\citep[e.g.][]{white94,Peterson04,bentz09,denney10},   in   order   to
measure robust lags in our dataset.  We measured cross-correlations of
H$\alpha$, H$\beta$ (on the red  and blue side spectra independently),
H$\gamma$ and \ion{He}{2} against the  $V$-band light curve from $-$20
to +40 days in increments of  0.25 days.  We show the CCF measurements
in    Figure~\ref{fig:ccf},    and     present    the    results    in
Table~\ref{table:cc_results}.      In    Table~\ref{table:cc_results},
\tpeak\ corresponds to  the peak of the CCF and  \tcen\ corresponds to
the centroid of  the CCF for all  points above 80\% of  the peak value
\citep{Peterson04}.

First, the blue and red side H$\beta$ lags are in excellent agreement,
with $\tcen  = 7.52^{+1.43}_{-1.06}$ and  8.59$^{+0.92}_{-1.21}$ days,
respectively, lending  confidence to  our data reduction  and analysis
techniques.  The H$\gamma$ lag of $\tcen=7.40^{+1.50}_{-1.32}$ days is
consistent with  the H$\beta$ measurements, while  the \ion{He}{2} lag
is consistent with zero  ($\tcen=1.83^{+4.15}_{-2.58}$ days).  We also
measure an H$\alpha$ lag, $\tcen= 13.89^{+1.39}_{-1.41}$ days, that is
substantially larger than that of H$\beta$.  The relative lags between
each line are consistent with  that seen in other reverberation mapped
systems,    along    with    previous    measurements    of    Arp~151
\citep[e.g.][]{bentz10lags}.  While the  2015 LCOGT reverberation lags
are a  factor of $\sim$2 larger  than the lags previously  measured by
\citet{bentz10lags} from  the 2008  Lick campaign,  Arp~151's H$\beta$
velocity width is  smaller than measured in the 2008  data, leading to
virtually the same virial products -- see \S~\ref{sec:mbh}.

\begin{figure*}
\begin{center}
\mbox{ \epsfysize=10.0cm \epsfbox{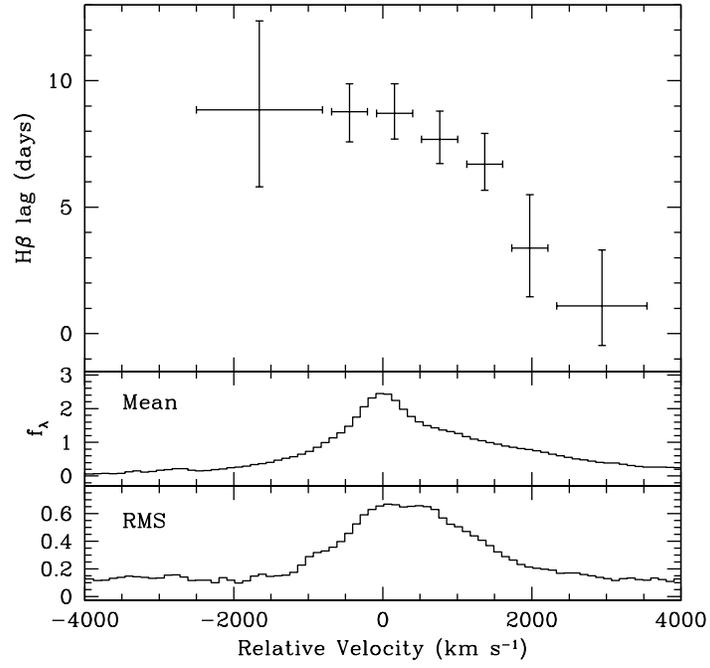}} 
\caption{Velocity-resolved   H$\beta$  lags,   as  derived   from  our
  blue-side FLOYDS  data.  The  upper panel shows  the lag  (using the
  $\tau_{\rm  cen}$   measure)  for  each  velocity   bin,  where  the
  horizontal error  bar represents  the bin  size.  The  bottom panels
  show the mean and rms continuum-subtacted spectra.  }
 \label{fig:v_ccf}
\end{center}
\end{figure*}

Past data  on Arp~151 have demonstrated  velocity-resolved variability
and  lags   \citep{Bentz08,bentz10memecho,pancoast14b}.   We  measured
H$\beta$ light curves  in seven velocity bins across the  width of the
line (utilizing our blue-side  FLOYDS data), and cross-correlated each
with  the $V$-band  light curve.   We plot  our velocity-resolved  lag
results  (\tcen)  in  Figure~\ref{fig:v_ccf}, which  shows  a  similar
morphology  as  previous  velocity-resolved   work  in  Arp~151.   The
H$\beta$  lag response  is  clearly asymmetric,  with  an $\sim$8  day
difference between  the lag in the  H$\beta$ core and the  red wing of
the line. The lag profile is similar in shape to that seen in the 2008
Lick  data  \citep{Bentz08,bentz10memecho},   despite  the  factor  of
$\sim2$ increase in the overall  H$\beta$ lag since 2008.  A symmetric
H$\beta$ lag response, declining on both the blue and red wings of the
line, would be  consistent with the BLR clouds  in Keplerian, circular
orbits around  the black hole \citep{Welsh91}.   However, the velocity
lags such as that seen in Arp~151,  with higher lags on the blue wing,
with  a decline  towards the  red, are  generally consistent  with BLR
models   with  radial   inflow   and/or  a   warped  disk   morphology
\citep[e.g.][]{bentz10memecho,pancoast14a}.

\section{Line Widths and Black Hole Mass Estimate}\label{sec:mbh}

The width of  the broad H$\beta$ line is needed  for determining black
hole masses  from reverberation data,  which we measure  following the
procedure  of  \citet{Barth15}.   First  we use  the  results  of  the
spectral decompositions of the nightly  second order FLOYDS spectra to
remove all components except for  the broad H$\beta$ emission from the
data.  From these  broad H$\beta$-only spectra, we  construct mean and
rms  spectra  and measure  the  line  width parameters  following  the
methods  of \citet{Peterson04}.   Uncertainties are  calculated via  a
Monte  Carlo resampling  technique.   The  instrumental resolution  of
FLOYDS with  the 1\farcs6 slit near  the H$\beta$ line is  FWHM=7 \AA,
corresponding  to $\approx$430  km s$^{-1}$  (or $\sigma_\mathrm{inst}
\approx$183 km s$^{-1}$),  and we subtract this  small contribution in
quadrature for  our final results.   The broad H$\beta$ from  the mean
spectrum  has  a  FWHM(H$\beta_{\rm  mean}$)=2872$\pm$90  km  s$^{-1}$
[$\sigma_\mathrm{line}$(H$\beta_{\rm mean}$)=1591$\pm$55 km~s$^{-1}$],
while for the rms  spectrum it is FWHM(H$\beta_{\rm rms}$)=1833$\pm$61
km~s$^{-1}$ [$\sigma_\mathrm{line}$(H$\beta_{\rm  rms}$)=879$\pm$57 km
  s$^{-1}$].

The   measured   H$\beta$   virial    product,   defined   as   VP   =
$\rblr$($\Delta$$V$)$^2$/$G$,      is     determined      by     using
$\Delta$$V$=$\sigma_\mathrm{line}$(H$\beta_{\rm       rms}$)       and
$\rblr$=$c$$\tau_\mathrm{cen}$  \citep[see][]{Peterson04},   which  we
find to be VP=1.13$^{+0.26}_{-0.22} \times$10$^{6}$~$M_{\odot}$.  This
value for the virial product is consistent to within the uncertainties
with     previous    determinations     of     \mbh\    in     Arp~151
\citep[e.g.][]{bentz10lags},     which     also     indicates     that
$\sigma_\mathrm{line}  \propto  \tau^{-0.5}$,  as is  expected  for  a
virial relationship between time lag and line width.

To convert our directly measured virial product into a black hole mass
requires an additional dimensionless factor, $f$, which depends on the
geometry, kinematics  and orientation of  the BLR.  While  recent work
has made great  progress in directly determining the value  of $f$ for
individual                                                         AGN
\citep{pancoast11,brewer11b,pancoast12,pancoast14a,pancoast14b},     we
adopt a value of $f$=5.5, which  was found by \citet{Onken04} to bring
the AGN $\mbh$--$\sigma_{*}$ relationship into agreement with the same
relationship    for     quiescent    galaxies.     We     then    find
$\mbh$=6.2$^{+1.4}_{-1.2}   \times$10$^{6}$~$M_{\odot}$,   where   the
quoted  uncertainty comes  only  from the  uncertainty  on the  virial
product and not on  the adopted value of $f$.  As  \mbh\ is simply the
virial product  multipled by  the $f$-factor, this  result is  also in
agreement  with   previous  work   on  Arp~151  and   consistent  with
\cite{brewer11b} and \cite{pancoast14b} within the uncertainties.

\section{Summary and Future Prospects}

We have presented the first  totally robotic AGN reverberation mapping
measurements for  the nearby Seyfert 1  galaxy Arp~151 as part  of the
LCOGT AGN Key Project.  Robust  lag measurements were measured for the
H$\beta$, H$\alpha$ and H$\gamma$ lines, and we were able to measure a
velocity-resolved H$\beta$ lag with a  profile similar to that seen in
previous  work.  By  combining the  lag measurements  with line  width
measurements,   we    infer   a    central   black   hole    mass   of
$\mbh$=6.2$^{+1.4}_{-1.2}   \times$10$^{6}$~$M_{\odot}$,    again   in
accordance with previous measurements in  this system.  Arp~151 is now
one of just  a few AGN where  it has been shown that  the virial product
remains  constant  against  substantial  changes  in  H$\beta$  lag  and
luminosity. NGC~5548 is the best example with many years of monitoring
and  an order  of magnitude  variation  in luminosity  over the  years
\citep{Bentz07}.

The reverberation  mapping portion  of the LCOGT  AGN Key  Project has
collected data  on other low-redshift  AGN as well as  higher redshift
quasars.  In  particular, we are  monitoring a sample of  moderate and
high redshift  objects with  a focus on  the Mg II  and C  IV emission
lines  and their  lags in  order  to determine  a direct  $\rblr$--$L$
relation for  these emission lines.   The robotic nature of  the LCOGT
network  and the  large  allocation of  telescope  resources are  both
critical for the success of the project.

\acknowledgments
Research by A.J.B. and L.P. are supported by NSF grant AST-1412693.
Research by DJS is supported by NSF grant AST-1412504 and AST-1517649.
Research by TT is supported by NSF grant AST-1412315 and a Packard Research Fellowship.
ERC and SC gratefully acknowledge the receipt of research grants from the National Research 
Foundation (NRF) of South Africa.
\bibliographystyle{apj}


\clearpage

\begin{deluxetable*}{lcccc}

\tablecolumns{5}
\tablecaption{Cross-correlation Lag Results \label{table:cc_results}}
\tablehead{
\colhead{Emission Line}  & \colhead{\fvar} & \colhead{CCF rmax} & \colhead{$\tau_{\rm cen}$} & \colhead{$\tau_{\rm peak}$}   \\
 & & & \colhead{(days)}& \colhead{(days)} 
 }
\\
\startdata
H$\alpha$ & 0.20$\pm$0.02 & 0.87 & 13.89$^{+1.39}_{-1.41}$ & 13.50$^{+4.00}_{-3.75}$\\
H$\beta$ blue-side & 0.29$\pm$0.03 & 0.93 & 7.52$^{+1.43}_{-1.06}$ &  5.75$^{+0.75}_{-1.00}$\\
H$\beta$ red-side & 0.32$\pm$0.03 & 0.93 & 8.59$^{+0.92}_{-1.21}$ &  6.00$^{+1.75}_{-1.25}$\\
H$\gamma$ & 0.30$\pm$0.03 & 0.91 & 7.40$^{+1.50}_{-1.32}$ &  5.75$^{+0.75}_{-1.25}$\\
\ion{He}{2} & 0.70$\pm$0.07 & 0.47 & 1.83$^{+4.15}_{-2.58}$ &  1.00$^{+4.75}_{-1.75}$
\enddata
\tablenotetext{}{All lags are measured relative to the $V$-band light curve, and are given in the observed frame. All line fluxes include constant narrow-line contributions, slightly decreasing the value of \fvar.}
\end{deluxetable*}

\end{document}